\documentstyle[aps,prl,twocolumn,floats,psfig]{revtex}

\begin{document}

\bibliographystyle{prsty}

\title{
Tunneling with dissipation and decoherence for a large spin.
\vspace{-1mm}}
\author{
E. M. Chudnovsky and  X. Martinez-Hidalgo }
\address{
Department of Physics and Astronomy, Lehman College, City University of New York, \\
250 Bedford Park Boulevard West, Bronx, New York 10468-1589. U.S.A. \\and\\
Departament de F\'{i}sica Fonamental, Universitat de Barcelona\\
Diagonal 647, 08028 Barcelona, Spain\\
\smallskip
{\rm(Received 11 January 2002)}
\bigskip\\
\parbox{14.2cm}
{\rm We present rigorous solution of problems of tunneling with
dissipation and decoherence for a spin of an atom or a molecule in
an isotropic solid matrix.
Our approach is based upon switching to a rotating coordinate
system coupled to the local crystal field.
We show that the spin of a molecule can be used in a qubit only if
the molecule is strongly coupled with its atomic environment.
This condition is a consequence of the conservation of the total
angular momentum (spin + matrix), that has been largely ignored in
previous studies of spin tunneling.
\smallskip
\begin{flushleft}
\end{flushleft}
}}
\maketitle

The problem of tunneling of a large spin
\cite{Korenblit,Rolf,Leo,CGB,Zaslavsky,Garanin,Loss1-Henley,Garg,book}
has received considerable attention lately in connection with
spin-10 magnetic molecules, Mn-12 and Fe-8,
\cite{Roberta,Villain,Jonathan,Zvezdin,GarChu,WenSes,Kent,Sarachik}.
High-spin molecules have been proposed as qubits for quantum
computers \cite{qubit}.
It is, therefore, important to understand the effect of the atomic
environment on spin tunneling and decoherence.
In this Letter we present rigorous solution of both problems for a
spin of a molecule in an isotropic solid.

The Caldeira-Leggett approach \cite{CL} to the problem of spin
tunneling with dissipation due to phonons was outlined in the
second of Refs. \cite{CGB}.
It has been applied in Ref. \cite{Garg-2}, though a general
expression for the effective action has not been obtained.
The correct formulation of the problem should account for the
conservation of the total angular momentum: spin ${\bf S}$ plus
the angular momentum ${\bf L}$ of the atomic lattice
\cite{ChuPhon}.
In the absence of the external field, tunneling of ${\bf S}$
between ${\Uparrow}$ and ${\Downarrow}$ should be accompanied by a
simultaneous co-flipping of ${\bf L}$ and is possible only if
${\bf J}={\bf S}+{\bf L}=0$.
This can be satisfied by an integer but not by a half-integer $S$,
which is another way to look at the Kramers theorem.
The necessity to have $L=S$ in the tunneling state results in the
mechanical rotational energy $E_{r}=({\hbar}S)^{2}/2{\cal{I}}$
where ${\cal{I}}\,{\sim}\,{\rho}l^{5}$ is the moment of inertia of
the atomic lattice; ${\rho}$ and $l$ being the mass density and
the linear dimension of the solid matrix containing ${\bf S}$.
This energy should be compared with the tunneling splitting
${\Delta}$.
The condition ${\Delta}\,>\,E_{r}$ is needed for the tunneling
state with $L=S$, $J=0$ to be the ground state of the system,
while in the opposite case of ${\Delta}\,<\,E_{r}$ the ground
state should be $L=0$, $J=S$, with ${\bf S}$ frozen along the
anisotropy axis.
This translates into a minimal size $l$ (typically of order 1-10
nanometers) of a free particle whose spin can tunnel between
equilibrium orientations.
%
%
%

The problem of the decoherence is even more subtle.
On one hand, the matrix elements of the conventional spin-lattice
interaction due to the electrostatic crystal field vanish between
$(|\Uparrow>+|\Downarrow>)$ and $(|\Uparrow>-|\Downarrow>)$
tunneling spin states \cite{Chu}.
On the other hand, the real-time coherent oscillations of ${\bf
S}$ in a solid matrix must be accompanied, through the
conservation of the angular momentum, by the oscillating shear
deformation of the solid.
This should be the source of decoherence.
The relevant size of the solid involved in that process is
$l_{c}=c_{t}/{\omega}_{c}$, where $c_{t}$ is the velocity of the
transverse sound and ${\omega}_{c}={\Delta}/{\hbar}$ is the
oscillation frequency.
We shall see that
$\Gamma=E_{r}/{\hbar}\,{\sim}\,{\hbar}S^{2}/{\rho}l_{c}^{5}$
emerges as a frequency scale that determines the decoherence rate.

The imaginary-time action of the system consists of the action of
the spin in a crystal field of the atomic lattice and the action
of the lattice. The spin action can be written as
\begin{equation}
I_{s}[{\bf n}]=I_{WZ}[{\bf
n}]+\int_{0}^{{\hbar}{\beta}}d{\tau}E_{s}({\bf n})\;,
\end{equation}
where $I_{WZ}$ is the Wess-Zumino action \cite{book}, $E_{s}$ is
the energy of the magnetic anisotropy due to the crystal field,
${\bf n}(\tau)={\bf S}(\tau)/S$, and ${\beta}=1/T$.
The anisotropy is determined by the local atomic environment of
the atom or molecule with spin ${\bf S}$.
The global symmetry of the lattice is less important since the
wavelengths involved in the problem (see below) are always large
compared to the atomic scale.
For that reason, and in order to simplify calculations, it is
convenient to choose an isotropic (e.g., amorphous) solid that is
characterized by two elastic moduli only, ${\mu}$ and ${\lambda}$,
which determine velocities of the transverse and longitudinal
sound, $c_{t}=({\mu}/{\rho})^{1/2}$,
$c_{l}=[({\lambda}+2{\mu})/{\rho}]^{1/2}$.
Within linear elastic theory the corresponding phonon action is
\begin{equation}
I_{l}[{\bf u}]=\int_{0}^{{\hbar}{\beta}}d{\tau}{\int}d^{3}{\bf
r}\left\{\frac{1}{2}{\rho}{\dot{\bf
u}}^{2}+{\mu}(u_{ij})^{2}+\frac{\lambda}{2}u_{kk}^{2}\right\}\;,
\end{equation}
where ${\bf u}(\tau)$ is the phonon displacement field and
$u_{ij}=\frac{1}{2}({\partial}_{i}u_{j}+{\partial}_{j}u_{i})$ is
the strain tensor.

We shall assume throughout this Letter that the interaction of the
spin with its atomic environment is contained {\it entirely} in
the structure of the crystal field responsible for the magnetic
anisotropy.
For example, in Mn-12 the spin Hamiltonian is dominated by the
uniaxial anisotropy, ${\cal{H}}_{s}=-D({\bf e}{\cdot}{\bf S})^{2}$
where ${\bf e}$ is the unit vector along the anisotropy axes.
The conventional way to introduce the spin-phonon interaction
\cite{Villain,GarChu} is to consider a small perturbation of ${\bf
e}$ by phonons, ${\delta}{\bf e}=[{\delta}{\bf \phi}{\times}{\bf
e}]$, where ${\delta}{\bf \phi}=\frac{1}{2}{\nabla}{\times}{\bf
u}$ is the local rotation.
This results in the spin-phonon Hamiltonian
\begin{equation}
{\cal{H}}_{sp}=D\{S_{z},S_{x}\}{\omega}_{zx}+D\{S_{z},S_{y}\}{\omega}_{zy}\;,
\end{equation}
where $\{S_{i},S_{j}\}$ is the anti-commutator and
${\omega}_{ij}=\frac{1}{2}({\partial}_{i}u_{j}-{\partial}_{j}u_{i})$.
The effective spin action,
\begin{equation}
I_{eff}[{\bf n}]=I_{s}[{\bf n}]+I_{env}[{\bf n}]\;,
\end{equation}
should be obtained by computing the environmental action
$I_{env}$,
\begin{equation}
\exp\left(-\frac{I_{env}[{\bf
n}]}{\hbar}\right)=\frac{1}{Z_{l}}\oint{\cal{D}}{\bf
u}\exp\left(-\frac{I_{l}[{\bf u}]}{\hbar}-\frac{I_{int}[{\bf
n},{\bf u}]}{\hbar}\right)\;,
\end{equation}
where
\begin{equation}
Z_{l}=\oint{\cal{D}}{\bf u}({\bf
r},{\tau})\exp\left(-\frac{I_{l}[{\bf u}]}{\hbar}\right)\;.
\end{equation}
The conventional approach \cite{CGB,Garg-2} is to choose
$I_{int}=\int_{0}^{{\hbar}{\beta}}d{\tau}{\cal{H}}_{sp}({\bf n})$
with ${\cal{H}}_{sp}$ of Eq.(3) or similar.
Then the integration in Eq.(5), though cumbersome, is Gaussian
and, in principle, can be performed exactly.
As has been discussed above, the problem with such an approach is
that the tunneling of ${\bf S}$ from the ${\Uparrow}$ state at
${\tau}=-\infty$ to the ${\Downarrow}$ state at ${\tau}=+\infty$
must formally involve the co-flipping of ${\bf L}$, that is, the
change of the mechanical rotation of the solid from clockwise to
counterclockwise.
In a fixed coordinate system the corresponding instanton involves
large dispacements ${\bf u}$ which are difficult to work with.

The difficulty mentioned above can be avoided by switching to a
coordinate system that is centered at the spin and is firmly
coupled to the local anisotropy axes. It rotates in the presence
of the time-dependent shear deformation.
In such a coordinate system the magnetic anisotropy $E_{s}[{\bf
n}]$ remains unaffected by phonons and the tunneling of the spin
is accompanied by small lattice displacements only.
The new term appears in the energy, though,
\begin{equation}
E'_{s}=-{\hbar}{\bf S}{\cdot}{\bf \Omega}\;,
\end{equation}
where ${\bf \Omega}={\delta}\dot{{\bf
\phi}}=\frac{1}{2}{\nabla}{\times}\dot{{\bf u}}$.
This can be considered as the consequence of the fact that
rotation is equivalent to the magnetic field.
In the rotating coordinate system $I_{int}$ in Eq.(5) becomes
\begin{equation}
I_{int}=-i{\hbar}S\int_{0}^{{\hbar}{\beta}}d{\tau}{\dot{\bf
n}}\left[\frac{1}{2}{\nabla}{\times}{\bf u}\right]_{{\bf r}=0}\;,
\end{equation}
where we have integrated by parts to move the time derivative to
${\bf n}$.
Consequently, Eq.(5) is again a simple Gaussian integral on phonon
variables.

Equations (7) and (8) can be mistakenly taken for a parameter-free
spin-phonon interaction \cite{Villain}.
It is important to understand, therefore, that they are solely the
consequence of switching to a rotating coordinate system, where
the interaction between ${\bf S}$ and the lattice is contained, to
all orders on ${\bf u}$, in the anisotropy energy $E_{s}[{\bf
n}]$.
To illustrate the equivalence of this approach to the conventional
method of studies of the spin-phonon interaction, let us compute,
e.g., the width, ${\Gamma}_{1}$, of the $T=0$ spin-precession
resonance for ${\cal{H}}_{s}=-DS_{z}^{2}$, that is the rate of the
one-phonon decay $|m=S-1>{\rightarrow}\;|m=S>$, provided by Eq.(3)
and by Eq.(7).
After standard quantization of ${\bf u}$ (see below for details)
Eq.(3) gives \cite{Orbach,Villain,GarChu}
${\Gamma}_{1}=S(2S-1)^{2}D^{2}{\omega}_{1}^{3}/12{\pi}{\hbar}{\rho}c_{t}^{5}$,
while Eq.(7) gives
${\Gamma}_{1}'=S{\hbar}{\omega}_{1}^{5}/12{\pi}{\rho}c_{t}^{5}$,
where ${\hbar}{\omega}_{1}$ is the distance between the $m=S$ and
$m=S-1$ levels.
Observing that this distance equals $(2S-1)D$, one immediately
obtains ${\Gamma}_{1}={\Gamma}_{1}'$.
The equivalence of the two methods in general can be traced to the
equation of motion for the spin operator, ${\hbar}(d{\bf
S}/d{\tau})=[{\cal{H}},{\bf S}]$.

We shall now proceed to the computation of the Caldeira-Leggett
action.
In terms of phonon modes,
\begin{equation}
{\bf u}({\bf r},{\tau})=\frac{1}{\sqrt{N}}\sum_{{\bf
k}{\lambda}}e^{i{\bf k}{\cdot}{\bf r}}{\bf e}_{{\bf
k}{\lambda}}u_{{\bf k},{\lambda}}({\bf r},{\tau})\;,
\end{equation}
$I_{l}$ and $I_{int}$ become
\begin{eqnarray}
I_{l}[{\bf u}] & = & \int_{0}^{{\hbar}{\beta}}d{\tau}\sum_{{\bf
k}{\lambda}}\frac{1}{2}M(\dot{u}_{-{\bf k}{\lambda}}\dot{u}_{{\bf
k}{\lambda}}+{\omega}^{2}_{{\bf k}{\lambda}}u_{-{\bf
k},{\lambda}}u_{{\bf k}{\lambda}}) \nonumber \\
I_{int}[{\bf u},{\bf n}] & = &
-i\int_{0}^{{\hbar}{\beta}}d{\tau}\sum_{{\bf k}{\lambda}}\dot{{\bf
n}}{\cdot}{\bf C}_{{\bf k}{\lambda}}u_{{\bf k}{\lambda}}\;,
\end{eqnarray}
where
\begin{equation}
{\bf C}_{{\bf k}{\lambda}}=\frac{1}{2\sqrt{N}}{\hbar}S(i{\bf
k}){\times}{\bf e}_{{\bf k}{\lambda}}\;,
\end{equation}
${\bf e}_{{\bf k},{\lambda}}$ are phonon polarization vectors,
${\bf e}_{-{\bf k},{\lambda}}{\cdot}{\bf e}_{{\bf
k}{\lambda}'}={\delta}_{{\lambda}{\lambda}'}\,$, $M$ is the unit
cell mass, and $N$ is the number of unit cells in the lattice.
The result of the integration over phonon variables in Eq.(5) can
be presented as a sum of the local and nonlocal terms,
\begin{equation}
I_{env}[{\bf n}]=I_{env}^{l}[{\bf n}]+I_{env}^{nl}[{\bf n}]\;,
\end{equation}
where the local part is given by
\begin{equation}
I_{env}^{l}[{\bf n}] =
\frac{{\hbar}^{2}S^{2}}{12{\mu}V_{c}}\int_{0}^{{\hbar}{\beta}}d{\tau}{\dot{\bf
n}}^{2}\;,
\end{equation}
and the nonlocal part is
\begin{eqnarray}
I_{env}^{nl} & = &
-\frac{9{\hbar}^{2}S^{2}}{32{\pi}{\mu}c_{t}^{3}}\int_{0}^{{\hbar}{\beta}}d{\tau}
\nonumber \\ &  & \int_{-\infty}^{\infty}
d{\tau}'f({\omega}_{D}^{t}|{\tau}-{\tau}'|) \frac{[{\dot{\bf
n}}({\tau})-{\dot{\bf n}}({\tau}')]^{2}}{|{\tau}-{\tau}'|^{4}} \;.
\end{eqnarray}
Here $V_{c}=M/{\rho}$ is the unit cell volume,
${\omega}_{D}^{t}=c_{t}/V_{c}^{1/3}$ is the Debye frequency for
the transverse phonon branch, and
\begin{equation}
f(x)=\frac{1}{6}\int_{0}^{x}dx'{x'}^{3}e^{-x'}\;.
\end{equation}

The typical scale of the time derivative in Eqs. (13) and (14) is
set by the instanton frequency, ${\omega}_{i}$, which also
determines the temperature, $T_{c}{\sim}{\hbar}{\omega}_{i}$, of
the crossover from superparamagnetism to quantum tunneling of
${\bf S}$. The scale of $T_{c}$ is set by $E_{s}/S$ \cite{book}.
It is easy to see that
$|I_{env}^{nl}/I_{env}^{l}|\,{\sim}\,(T_{c}/T_{D})^{2}$. This
ratio is small as long as the energy of the magnetic anisotropy is
small compared to the elastic energy, which is normally the case.
The relative effect of the atomic environment on spin tunneling is
given by the ratio $I_{env}/I_{s}$, which is of the order of
$E_{s}/{\mu}V_{c}$. For a molecule of spin ${\bf S}$, {\it
rigidly} imbedded in a solid matrix, this ratio is typically
$10^{-6}-10^{-3}$. On the contrary, {\it a loose} coupling of the
magnetic molecule with its atomic environment is equivalent to
${\mu}\,{\rightarrow}\,0$, which must result in a drastic increase
of $I_{eff}$ and the corresponding exponential decrease of the
tunneling rate.

The advantage of our method becomes apparent when one wants to
compute the rate of the decoherence for quantum oscillations of
${\bf S}$ between ${\Uparrow}$ and ${\Downarrow}$. Here we assume
that the tunneling splitting $\Delta$ has been already
renormalized by the effects studied above. If $\Delta$ is small,
the problem can be trancated to the spin 1/2 problem with only two
states: the ground state
$|0>=\frac{1}{\sqrt{2}}(|\,{\Uparrow}>+\;|\,{\Downarrow}>)$ and
the excited state,
$|1>=\frac{1}{\sqrt{2}}(|\,{\Uparrow}>-\,|\,{\Downarrow}>)$. At
$T=0$ the decoherence occurs due to the transition
$|1>\,{\rightarrow}\,|0>$ via spontaneous emission of a phonon of
energy ${\Delta}={\hbar}{\omega}_{c}$. Such transitions seem to be
absent for the conventional magneto-elastic coupling that one
obtains by considering the {\it adiabatic} rotation of the
electrostatic crystal field (see, e.g., Eq.(3)).
To compute the decoherence rate, one has to take into account the
{\it dynamical} nature of the phonon \cite{Vleck}. This can be
done by switching to a rotating coordinate system coupled to the
local crystal field.

The general formula for the transition rate is
\begin{equation}
\Gamma = \frac{2\pi}{\hbar} \sum_{i\neq j} <i|\hat{V}|j>
  <j|\hat{V}|i> \delta(E_i - E_j) \;.
\end{equation}
Substituting here $\hat{V}=-{\hbar}\hat{\bf S}{\cdot}\hat{\bf
\Omega}$, we get
\begin{equation}
\Gamma=2{\hbar}<0|\hat{\bf S}|1>[J({\Delta})]<1|\hat{\bf S}|0>\;,
\end{equation}
where $J(\Delta)$ is the spectral function of the environmental
coupling,
\begin{equation}
J(\Delta)={\pi}\sum_{{\bf k}{\lambda}}<{\bf k}{\lambda}|\hat{\bf
\Omega}|0><0|\hat{\bf \Omega}|{\bf
k}{\lambda}>\delta(\Delta-{\hbar}{\omega}_{{\bf k}{\lambda}})\;.
\end{equation}
Computing the spin matrix elements, one gets
\begin{equation}
\Gamma=2{\hbar}S^{2}J(\Delta)\;.
\end{equation}

To compute the spectral function, let us consider
\begin{equation}
{\cal{J}}(\Delta)={\pi}\sum_{{\bf k}{\lambda}}{{\bf
\Omega}}^{*}_{{\bf k}{\lambda}}{\otimes}\,{{\bf \Omega}}_{{\bf
k}{\lambda}}\delta(\Delta-{\hbar}{\omega}_{{\bf k}{\lambda}})\;,
\end{equation}
where
\begin{equation}
{\bf \Omega}_{{\bf k}{\lambda}}=<0|\hat{\bf \Omega}|{\bf
k}{\lambda}>\;.
\end{equation}
Here
\begin{equation}
\hat{\bf \Omega}=\frac{1}{2}[{\nabla}{\times}\dot{\bf
u}]_{r=0}=\frac{1}{2{\rho}}[{\nabla}{\times}\hat{\cal{\bf
\Pi}}({\bf r},t)]_{r=o}
\end{equation}
and $\hat{\cal{\bf \Pi}}({\bf r},t)$ is the momentum of phonons in
terms of the operators of creation and annihilation,
\begin{equation}
\hat{\cal{\bf \Pi}}({\bf r},t)=-\frac{i}{\sqrt{V}}\sum_{{\bf
k}{\lambda}}\sqrt{\frac{{\hbar}{\omega}_{{\bf
k}{\lambda}}}{2{\rho}}}[\hat{a}_{{\bf k}{\lambda}}{\bf e}_{{\bf
k}{\lambda}}e^{i{\bf k}{\cdot}{\bf r}}-h.c.]\;.
\end{equation}
This gives
\begin{equation}
{\bf \Omega}_{{\bf
k}{\lambda}}=-i\sqrt{\frac{{\hbar}{\omega}_{{\bf
k}{\lambda}}}{2M}}{\bf c}_{{\bf k}{\lambda}}\;,\;\;\;{\bf c}_{{\bf
k}{\lambda}}=\frac{1}{2\sqrt{N}}i{\bf k}{\times}{\bf e}_{{\bf
k}{\lambda}}\;.
\end{equation}
Consequently, Eq.(20) becomes
\begin{equation}
{\cal{J}}(\Delta)=\frac{\pi}{2}\sum_{{\bf
k}{\lambda}}\frac{{\Delta}}{M}{\bf c}^{*}_{{\bf
k}{\lambda}}{\otimes}\,{\bf c}_{{\bf
k}{\lambda}}\delta({\Delta}-{\hbar}{\omega}_{{\bf k}{\lambda}})\;.
\end{equation}
The spectral function is $J=\frac{1}{3}Tr[{\cal{J}}(\Delta)]$,
which gives
\begin{equation}
J(\Delta)=\frac{\pi}{6}\sum_{{\bf
k}{\lambda}}\left(\frac{\Delta}{M}\right)\left(\frac{k^{2}}{4N}\right){\delta}_{{\lambda}t}
\delta({\Delta}-{\hbar}{\omega}_{{\bf k}{\lambda}})\;.
\end{equation}
The summation over phonon modes yields
\begin{equation}
J(\Delta)=\frac{{\Delta}^{5}}{24{\pi}{\hbar}^{5}{\rho}c_{t}^{5}}\;.
\end{equation}

Substituting Eq.(27) into Eq.(19) we finally obtain
\begin{equation}
\Gamma=\frac{{\hbar}S^{2}}{12{\pi}{\rho}l_{c}^{5}}
\end{equation}
for the decoherence rate.
Here we have introduced $l_{c}=c_{t}/{\omega}_{c}$.
The generalization to finite temperatures is trivial.
It results in the multiplication of the zero-temperature
${\Gamma}$ of Eq.(28) by the factor $\coth({\Delta}/2T)$.
Note that the form of Eq.(28) is independent of the form of the
spin Hamiltonian as long as the crystal field is the only source
of the spin-lattice coupling.
The dependence of $\Gamma$ on the parameters of the crystal field
is absorbed into ${\Delta}={\hbar}{\omega}_{c}$.

The physics behind Eq.(28) is this \cite{squid}.
The real-time quantum oscillations of ${\bf S}$ produce
oscillating torque in the surrounding matter.
That torque results in the torsion oscillations of the solid
adjacent to the magnetic molecule, so that ${\bf J}={\bf S}+{\bf
L}=0$.
The corresponding time-dependent elastic deformation is a coherent
quantum superposition of diverging and converging sound waves.
That deformation is confined within distances of order
$l_{c}=c_{t}/{\omega}_{c}$ from ${\bf S}$.
By order of magnitude, $l_{c}$ determines how far the sound goes
away from the spin during the period of coherent quantum
oscillations.
At times of order ${\Gamma}^{-1}$ the coherent superposition of
sound waves is destroyed by a diverging wave of energy $\Delta$
that escapes to the surface of the solid.
The energy ${\hbar}{\Gamma}$ determines the width of the excited
state $|1>$.
It is easy to see that
${\hbar}{\Gamma}\,{\sim}\,L^{2}/2{\cal{I}}_{c}$, where
$L={\hbar}S$ is the angular momentum associated with the spin and
${\cal{I}}_{c}\,{\sim}\,{\rho}l_{c}^{5}$ is the moment of inertia
of the volume of spatial dimensions $l_{c}$.

If the magnetic molecule is to serve as a qubit, the frequency
${\omega}_{c}$ should be sufficiently high, in the ballpark of
$10^{10}-10^{11}s^{-1}$.
Our study shows that it is possible only if the molecule is
rigidly coupled to a solid matrix.
This is contrary to the widely accepted view that the decoupling
from the environment should be beneficial for the work of the
qubit.
In fact, {\it any effort to make a loose connection of the spin
with its atomic environment, by e.g. having a magnetic molecule on
a some kind of a molecular leg, will most certainly decrease the
tunneling spitting} ${\Delta}$.

The rigid coupling to the atomic environment is also necessary to
provide a small decoherence rate.
According to Eq.(28) {\it the softening of the solid matrix with
respect to shear deformations, that results, e.g., in the decrease
of $c_t$ by a factor of four, increases $\Gamma$ by a factor of
one thousand}.
For $S=10$ and $c_{t}\,{\sim}\,10^{5}cm/s$, one obtains from
Eq.(28) ${\Gamma}\,{\sim}\,10^{-2}s^{-1}$ at
${\omega}_{c}\,{\sim}\,10^{10}s^{-1}$ and
${\Gamma}\,{\sim}\,10^{3}s^{-1}$ at
${\omega}_{c}\,{\sim}\,10^{11}s^{-1}$.
These numbers show that magnetic entities can, in fact, be
promising candidates for qubits.
It should be emphasized that Eq.(28) establishes the lower bound
on the decoherence rate.
Other effects unaccounted here, like interaction with nuclear
spins, with free electrons, etc., can bring the decoherence rate
up.

We thank Dmitry Garanin and Tony Leggett for helpful comments.
Discussions with Matthew Fisher and visiting members of ITP-USCB
during August 2001 are gratefully acknowledged.
This work has been supported by the NSF Grant No. 9978882.
Research of XMH has been funded by the Catalan Government.

\vspace{-0.5cm}


\end{document}